\documentclass[twocolumn,prl,aps,superbib,tightenlines,floatfix,superscriptaddress,showpacs]{revtex4}
\usepackage{amsmath}
\usepackage{graphicx}
\usepackage{amssymb}
\usepackage{color,soul}
\usepackage[T1]{fontenc}
\usepackage{ae,aecompl}

\begin{document}

\title{Kinks and antikinks of buckled graphene: A testing ground for $\phi^4$ field model}
\author{R. D. Yamaletdinov}
\affiliation{Nikolaev Institute of Inorganic Chemistry SB RAS, Novosibirsk 630090, Russia}
\author{V. A. Slipko}
\affiliation{Department of Physics and Technology, V. N. Karazin Kharkov National University, Kharkov 61022, Ukraine}
\affiliation{Institute of Physics, Opole University, Opole 45-052, Poland}
\author{Y. V. Pershin}
\affiliation{Department of Physics and Astronomy, University of South Carolina, Columbia, South Carolina 29208, USA}
\affiliation{Nikolaev Institute of Inorganic Chemistry SB RAS, Novosibirsk 630090, Russia}
\email{pershin@physics.sc.edu}


\begin{abstract}
Kinks and antikinks of the classical $\phi^4$ field model are topological solutions connecting its two distinct ground states.
 Here we establish an analogy between the excitations
of a long graphene nanoribbon buckled in the transverse direction and $\phi^4$ model results. Using molecular dynamics simulations,
we investigated the dynamics of a buckled graphene nanoribbon with a single kink and with a kink-antikink pair.
Several features of $\phi^4$ model have been observed including the kink-antikink capture
at low energies, kink-antikink reflection at high energies, and a bounce resonance. Our results pave the way towards
the experimental observation of a rich variety of $\phi^4$ model predictions based on graphene.
\end{abstract}
\maketitle   

Currently, there is strong interest in suspended graphene (the graphene above a trench) from both fundamental and application points of view.
Fundamentally, the advantages of graphene suspension include the elimination of substrate-induced carrier scattering, dopants and phonon leakage.
This provides an access to intrinsic properties of graphene such as
the intrinsic carrier  mobility~\cite{Bolotin08a},  mechanical  strength~\cite{lee2008}
and thermal conductivity~\cite{Balandin08a}. From the application point of view, multiple
possible applications of suspended graphene have been proposed, among them are electromechanical resonators~\cite{bunch2007}, electromechanical switchers~\cite{Sun14a}, nano-cantilever sensors~\cite{Lachut13a}, piezoresistive pressure sensors~\cite{Smith13a}, capacitive pressure sensors~\cite{chen2016ultra}, and capacitors with memory~\cite{pershin11c,Sedelnikova2016} (memcapacitors~\cite{diventra09a})~\footnote{Memcapacitors are two-terminal capacitive devices that can store and process~\cite{pershin15a} information on the same physical platform.}.

\begin{figure}[b]%
\includegraphics[width=80mm]{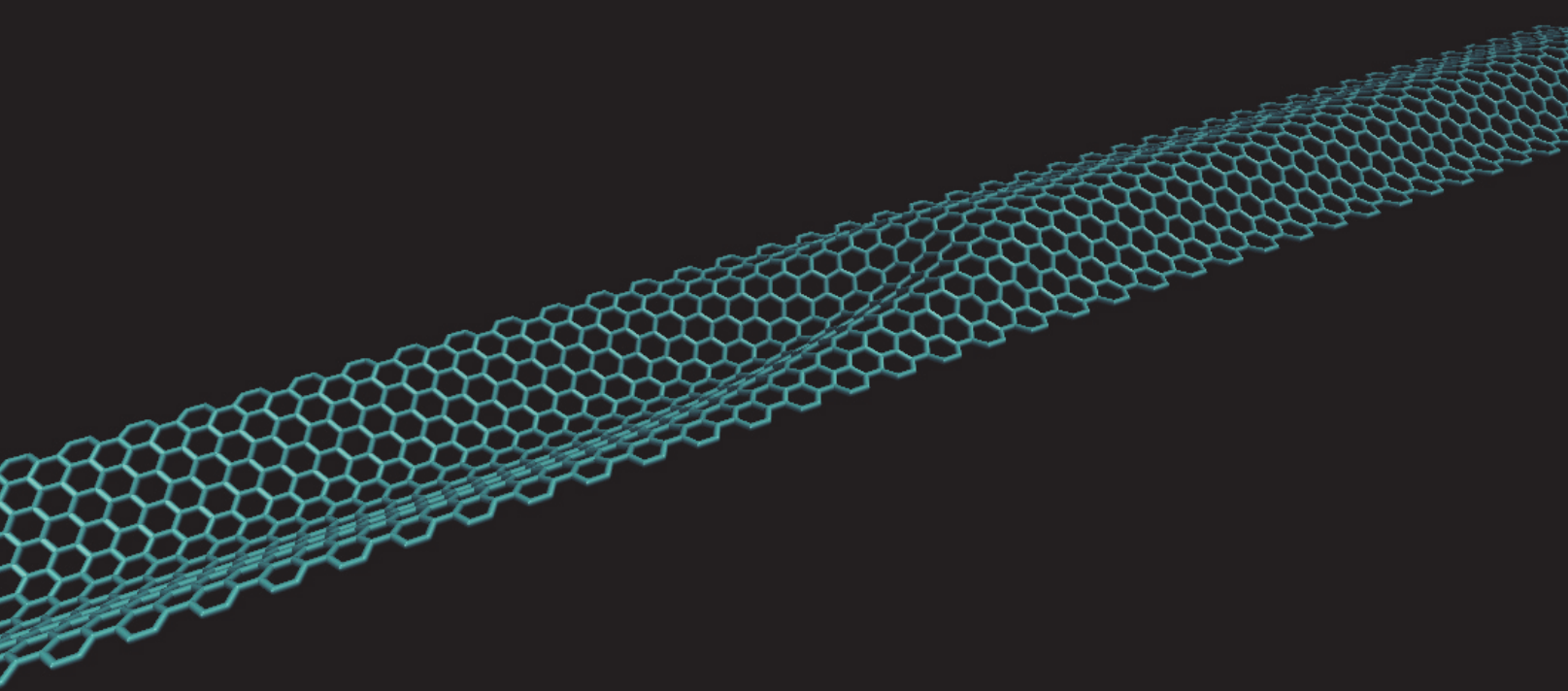}
\caption{Graphene kink.}
\label{fig1}
\end{figure}

Here, we consider a buckled graphene over a trench in a non-standard geometry in which the trench length is much longer than its width.
It is assumed that the graphene is relaxed in the direction along the trench and compressed in the transverse direction.
Using classical molecular dynamics (MD) simulations we demonstrate the existence of kinks (see Fig. \ref{fig1}) and antikinks in such buckled graphene -- the solutions connecting two minima of potential energy -- and investigate some of their properties.
To the best of our knowledge, this Letter is the first study of buckled graphene in such configuration.

Furthermore, we argue that such buckled graphene provides (to the certain extent) an experimental realization of the classical $\phi^4$ field model~\cite{rajaraman1982instantons,campbell1983,belova1997} in 1+1 dimensions, having applications in the areas of ferroelectrics~\cite{schmid1994multi,rowley2014ferroelectric,khare2006domain}, linear polymeric chains~\cite{RICE1980487}, quantum field theory~\cite{PhysRevD.10.4114,CAMPBELL1982297}, and even nuclear physics~\cite{PhysRevD.14.2093,PhysRevD.13.3398}.
This model is based on the dimensionless Lagrangian
\begin{equation}\label{eq:Lagrangian}
\mathcal{L}=\frac{1}{2}\int \textnormal{d}x \left[\left( \frac{\partial \phi}{\partial t}\right)^2-\left( \frac{\partial \phi}{\partial x}\right)^2
-\frac{1}{2}(1-\phi^2)^2\right]
\end{equation}
that leads to the Euler-Lagrange equation of motion of the form
\begin{equation}\label{eq:phi4}
  \frac{\partial^2 \phi}{\partial t^2}- \frac{\partial^2 \phi}{\partial x^2}= \phi-\phi^3.
\end{equation}
There are two stable constant solutions of Eq.~(\ref{eq:phi4}), $\phi=\pm 1$, and one unstable trivial,  $\phi=0$.
 Moreover, Eq.~(\ref{eq:phi4}) has a topologically non-trivial solutions
\begin{equation}\label{eq:Kink}
 \phi_K=\pm \tanh \left( \frac{x-Vt-x_0}{\sqrt{2(1-V^2)}}\right),
\end{equation}
where $\pm$ signs correspond to the kink and antikink, respectively, $V$ is the velocity, and $x_0$ is the position of the kink/antikink center at the initial
moment of time $t=0$.
In what follows, we present the details and results of our MD simulations and discuss the similarity between the graphene kinks (exemplified in Fig. \ref{fig1})
and the solutions (\ref{eq:Kink}) of Eq. (\ref{eq:phi4}).

\begin{figure}[t]%
(a)\includegraphics[width=70mm]{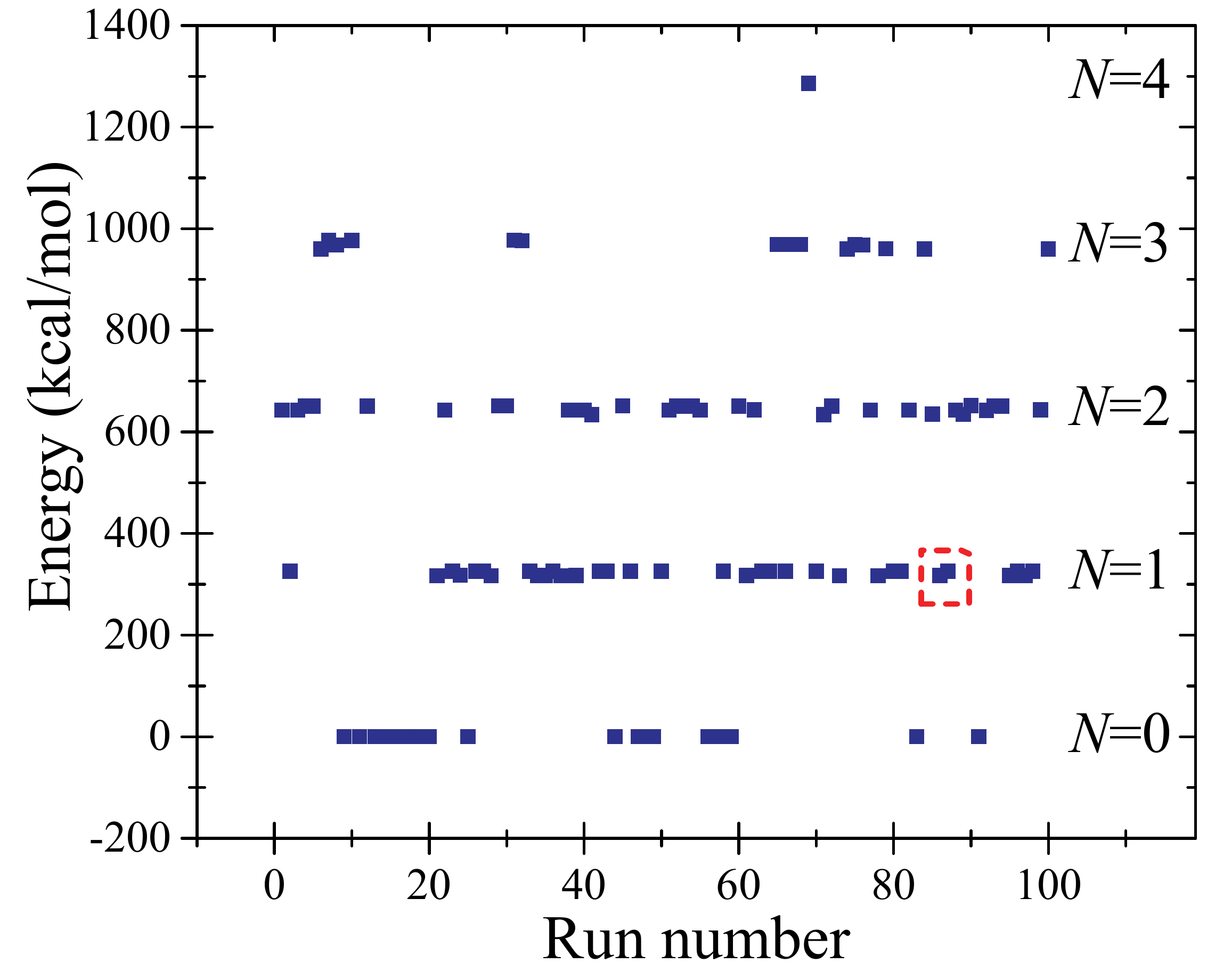} \\
(b) \hspace{0.25cm} \includegraphics[width=67mm]{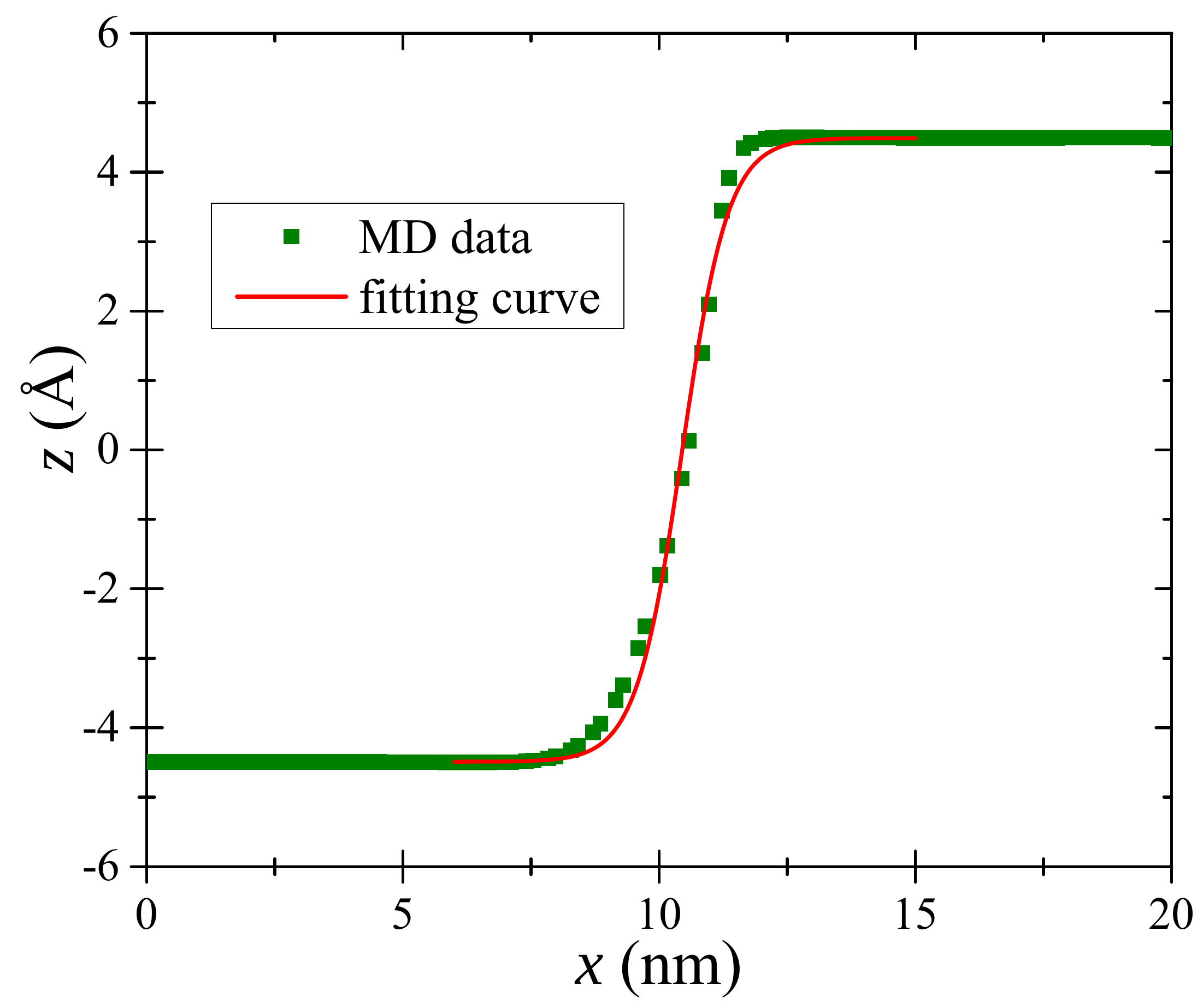}
\caption{(a) Energies of stable nanoribbon conformations found in
100 independent simulations of nanoribbon dynamics followed
by energy minimization. Zero energy corresponds to the nanoribbon
buckled up or down. The equidistant energy levels
correspond to the nanoribbons with different numbers of
kinks and antikinks $N$. (b) Geometry of the central chain of nanoribbon atoms in $\varepsilon_{1,0}$ kink found in the energy minimization. The fitting was made using Eq. (\ref{eq:Kink})-type curve.}
\label{fig2}
\end{figure}

MD simulations are frequently employed to model various physical aspects of graphene, carbone nanotubes and other nanoscale structures~\cite{Wang2009,Hu2009,Jiang2009,martins2010,Ni2010,Lebedeva2011,Ng2012,Kalosakas2013,Berdiyorov2014,Yamaletdinov17a}.
In this study, we used NAMD2~\cite{phillips05}, a
highly scalable massively parallel classical MD code~\footnote{NAMD was developed by the Theoretical and Computational Biophysics Group in the Beckman Institute for Advanced Science and Technology at the University of Illinois at Urbana-Champaign.}.
Specifically, we investigated the dynamics of a
graphene nanoribbon (membrane) of $L = 425$ \r{A} length and $w = 22$ \r{A}
width with clamped boundary conditions for the longer armchair
edges and free boundary conditions for the shorter edges
of the nanoribbon. The clamped boundary conditions were
implemented by fixing two first lines of carbon atoms of
longer edges. The buckling was realized by changing the
distance between the fixed sides from $w$ to $d < w$. For the
purpose of brevity, this Letter reports the results obtained
for a single value of $d/w = 0.9$.

\begin{figure*}[t]%
\hspace{0.5cm}
(a)
\includegraphics*[width=72mm]{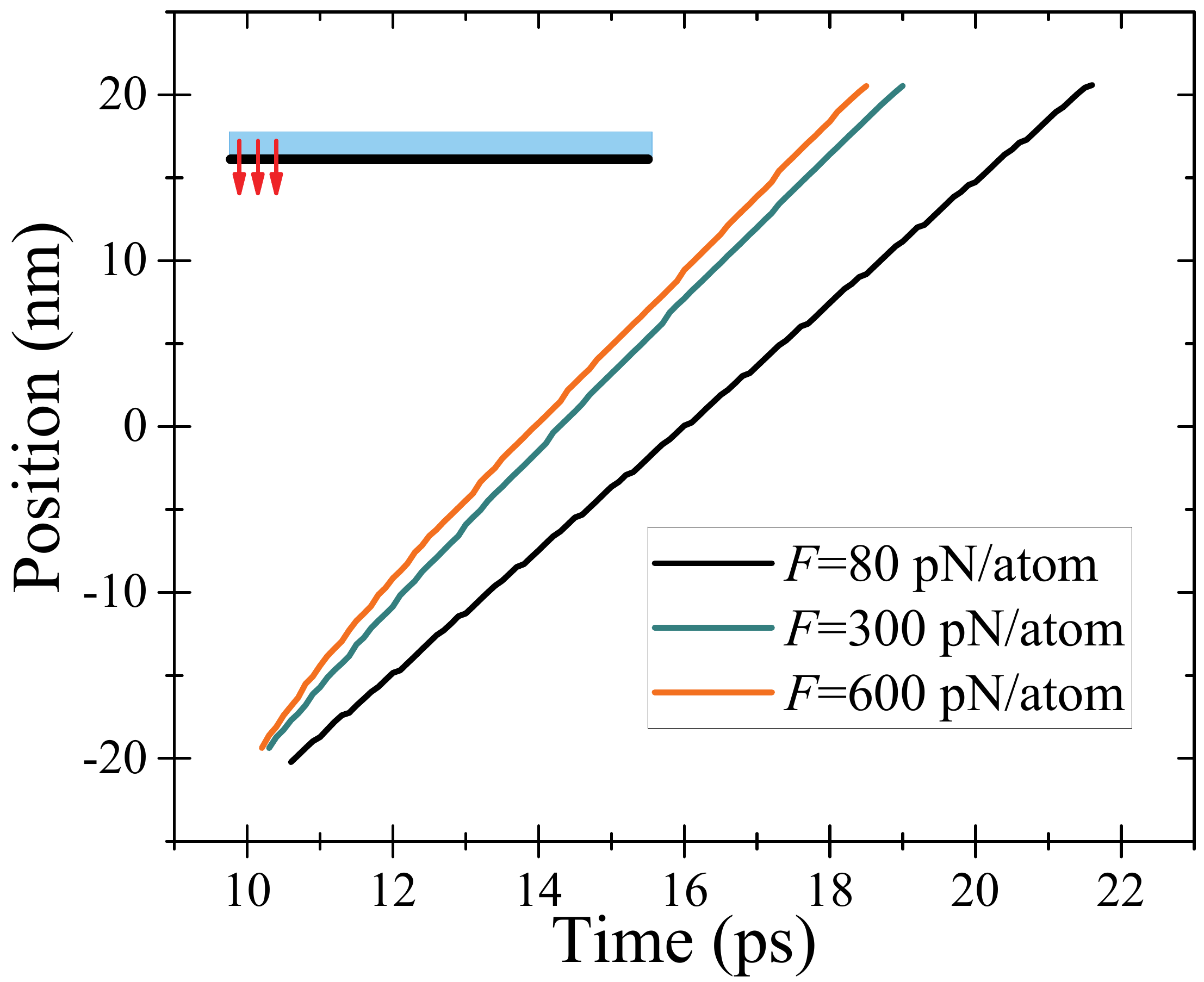}
\hspace{0.5cm}
(b)
\includegraphics*[width=70mm]{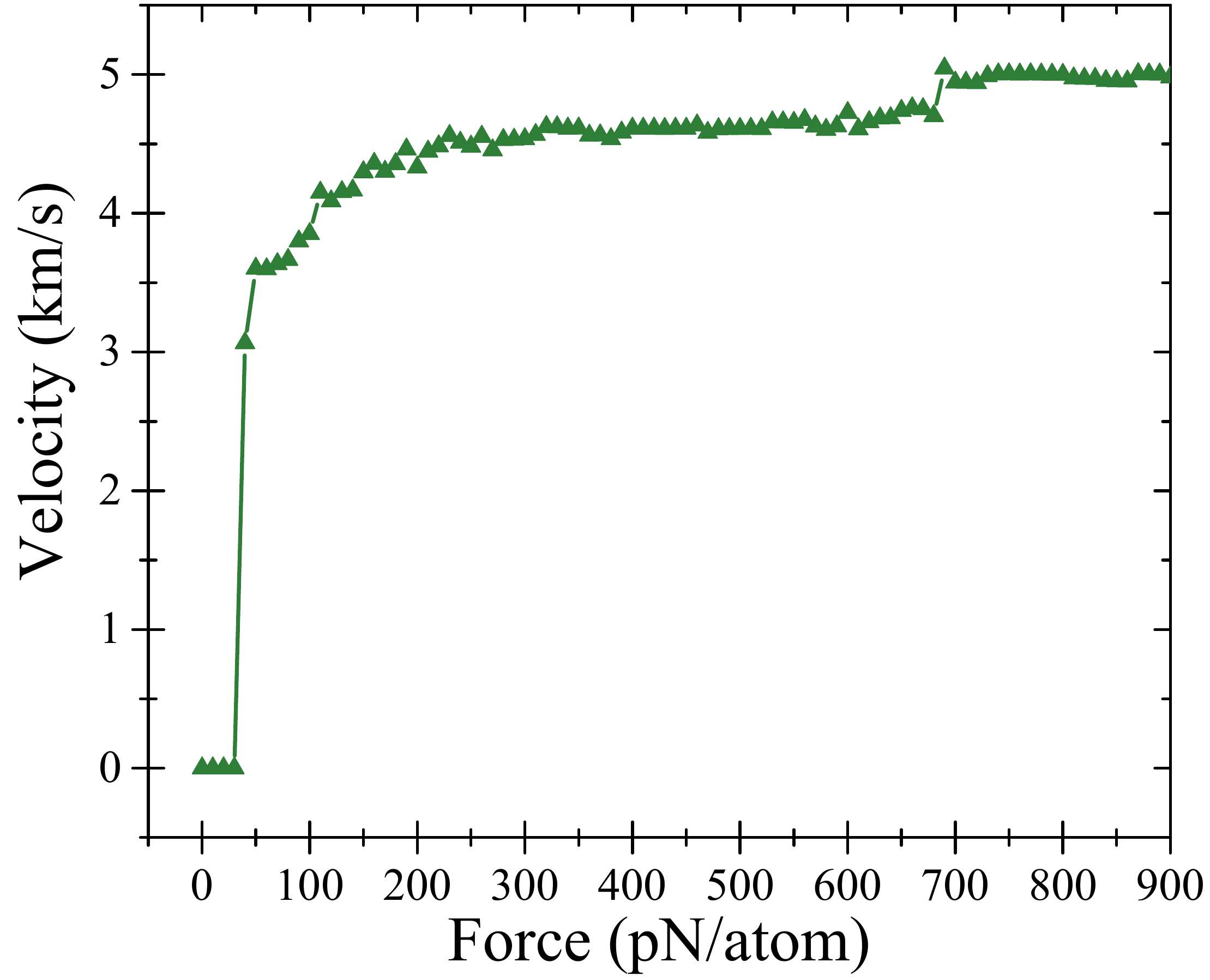}
\caption{(a) Kink position as a function of time for several values of the applied force. (b) Kink velocity as a function of the pulling force. Inset in (a): schematics of the generation of moving kink by locally applied forces (red arrows). Here, the thick (cyan) line represents the buckled membrane and thin (black) line - the membrane support surface.
}
\label{fig3}
\end{figure*}

In graphene, the interactions between the carbon atoms were described using the standard 2-body spring bond, 3-body angular bond (including the Urey-Bradley term), 4-body torsion angle and Lennard-Jones potential energy terms~\cite{JCC21367}. The interaction constants were optimized in order to fit the experimentally observed properties of graphene. The known values were selected for the equilibrium angles. The Lennard-Jones coefficients were choosen to match the AB stacking distance and energy of graphite~\cite{Chen2013}. A global optimization was performed over the remaining parameters to match the in-plane stiffness ($E_{2D}=342$ N/m), bending rigidity ($D=1.6$ eV) and equilibrium bond length ($a=1.421$ \r{A}) of graphene. We
performed a series of test calculations that have demonstrated
that the in-plane stiffness and bending rigidity of
a selected graphene nanoribbon are in perfect agreement
with the listed above values.

MD simulations were performed with 1 fs time step.
The van der Waals interactions were gradually cut off starting
at 10 \r{A} from the atom until reaching zero interaction 12
\r{A} away. For the sake of clarity, most of our molecular dynamics
calculations are made at zero temperature in terms of the Newtonian
dynamics.  In some of our calculations, kinks and antikinks were created
by applying external forces to groups of atoms near
the shorter edges of the nanoribbon (204 atoms in each
group). Typically, the external forces were applied for 10
ps and the nanoribbon dynamics was simulated for 30 ps from the moment
of force application.

Now, we consider the results of energy mimimization calculations. The purpose
of these calculations was to identify the energies
and shapes of individual kinks and antikinks as well as
to obtain the initial geometry for the kink/antikink dynamics calculations.
For this purpose, the Langevin dynamics of the
initially flat compressed nanoribbon was runned for 20
ps at $T = 293$ K using a Langevin damping parameter of
0.2 ps$^{-1}$ in the equations of motion. This simulation stage
was followed by 10000 steps of energy minimization.

Fig. \ref{fig2}(a) shows the energy of final shapes of the nanoribbon found in 100 runs
with identical initial conditions (the flat nanoribbon) and simulation parameters.
In consequence of the stochastic dynamics different numbers of kinks/antikinks are formed in different runs.
Fig. \ref{fig2}(a) demonstrates that the final energies form a set of
equidistant levels. We have explicitly verified that the each kink/antikink contributes approximately the same amount of energy into the total energy of final conformation.

Moreover, each of $N\geq 1$ energy levels in Fig. \ref{fig2}(a) has its unique fine structure. In particular, $N=1$ level is double-splitted (see two slightly different values of energy in the red dashed square in Fig. \ref{fig2}). The energies of these states are $\varepsilon_{1,0}=317.1$ kcal/mol and $\varepsilon_{1,1}=325.4$ kcal/mol. We have verified that all the energies in Fig. \ref{fig2}(a) are combinations of $\varepsilon_{1,0}$ and $\varepsilon_{1,1}$. Therefore, the level splitting increases with the number of kinks and antikinks $N$ as $N+1$.
In terms of geometry, the kink corresponding to $\varepsilon_{1,0}$ has a symmetric cross-section in the transverse to the trench direction.
The same plane cross-section of the kink corresponding to $\varepsilon_{1,1}$ is non-symmetric. Fig. \ref{fig2}(b) presents the profile of $\varepsilon_{1,0}$ kink along the trench. The fitting with $\phi^4$ model solution (Eq. \ref{eq:Kink}) reveals some small  asymmetry of this kink along the trench.

\begin{figure*}[htb]%
(a)
\includegraphics*[width=80mm]{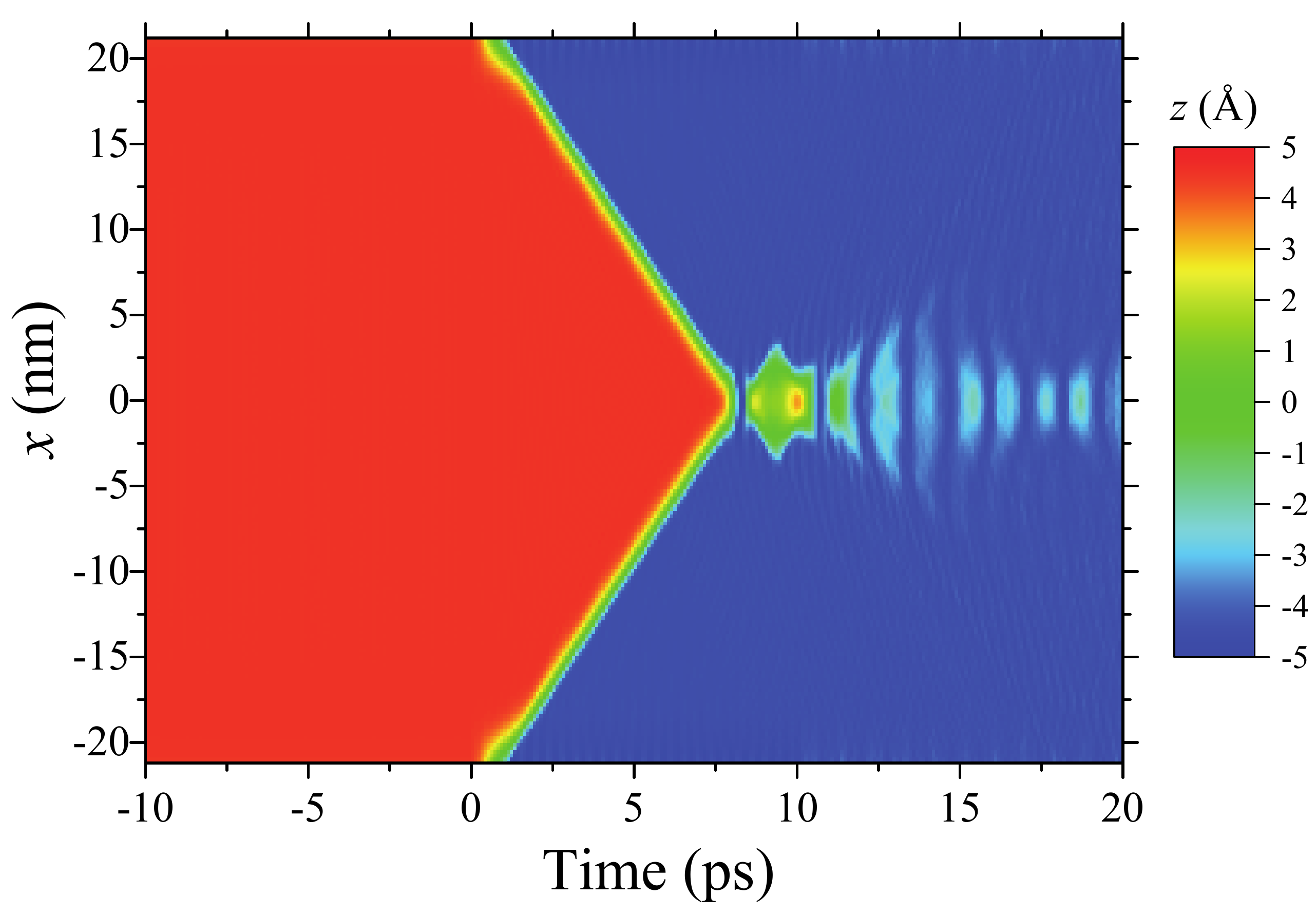}
(b)
\includegraphics*[width=80mm]{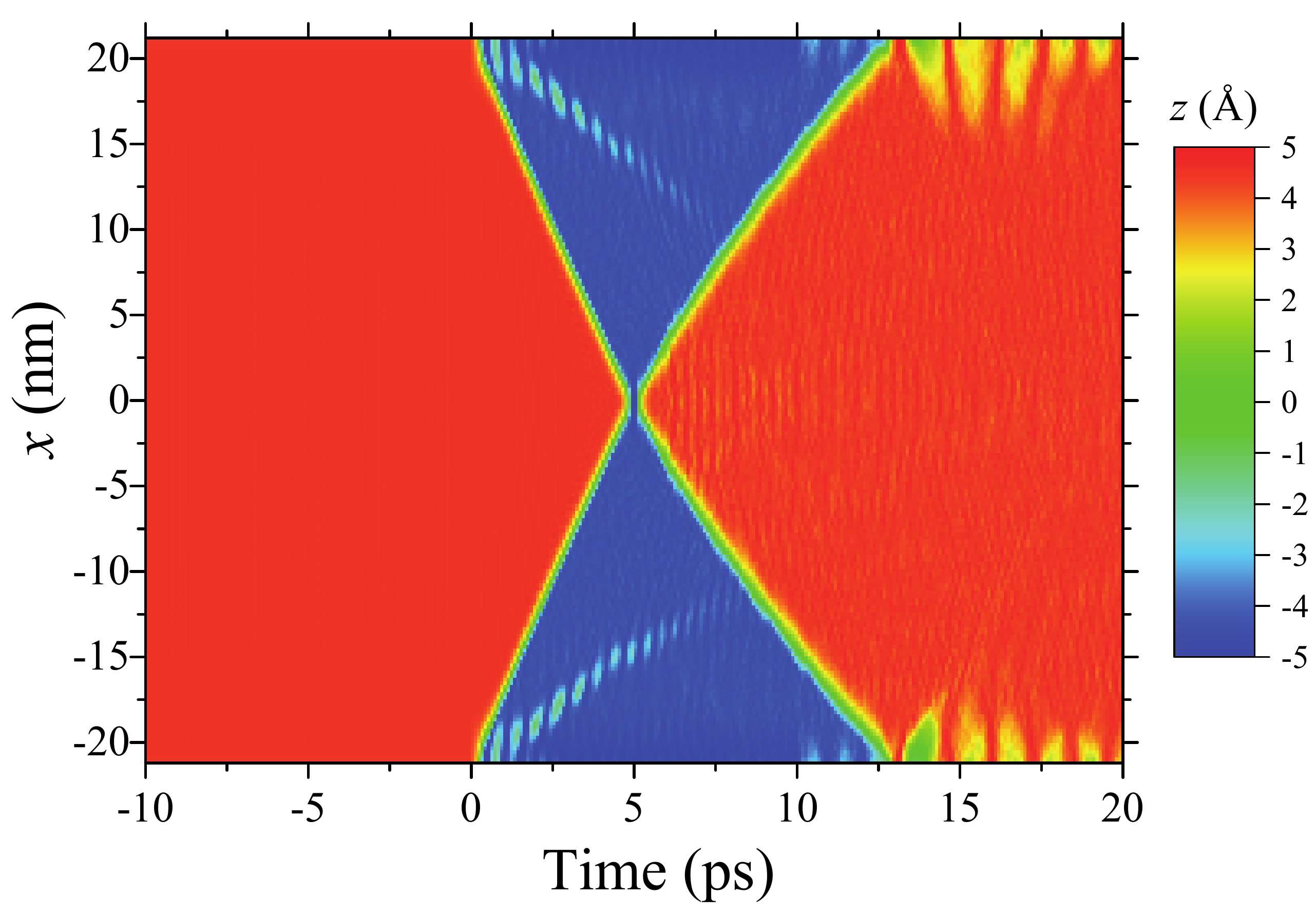}
\caption{Kink-antikink scattering: (a) kink and antikink are created by 60 pN/atom pulling force, and (b) kink and antikink are created by 200 pN/atom pulling force.
The plot was obtained using $z$-coordinates of the central chain of nanoribbon atoms (in the direction along the trench).
}
\label{fig4}
\end{figure*}

Next we consider the dynamics of a single kink propagating along the nanoribbon.
In these calculations, the optimized
buckled upwards nanoribbon was used as the initial
condition. The moving kink is created by pulling a group of atoms (204 atoms) near $x=-20$ nm boundary by a force
in $-z$ direction applied for 10 ps (see the inset in Fig. \ref{fig3}(a) for the numerical experiment schematics). According to the results of our MD simulations, the moving kinks are created at forces exceeding a threshold of $\sim 35$ pN/atom. Fig. \ref{fig3}(a) demonstrates that the position of kink changes linearly with time. Importantly, the slopes of Fig. \ref{fig3}(a) curves indicate the motion at a constant speed, which is higher for larger values of applied force.

It is thus tempting to use the applied force as a control parameter to select the desired velocity of kink.
Unfortunately, using this approach we were not able to generate slow-moving kinks. The data presented in Fig. \ref{fig3}(b) demonstrates that the slowest moving kink has the velocity of $\sim 3$ km/s and the velocity as a function of force saturates at about $5$ km/s. There are three characteristic speeds of sound in flat graphene (see, e.g., Ref.~\cite{Adamyan11a}): $c_{LA}=18.4$ km/s, $c_{TA}=16.5$ km/s, $c_{ZA}=9.2$ km/s. As the kink propagation is associated mainly with the out-of-plane deformation of graphene, $c_{ZA}$ provides the upper bound for the possible speed of kink propagation. Our results are in agreement with this bound.

According to numerical simulations of Eq. (\ref{eq:phi4}) (see, e.g., Ref.~\cite{goodman2005kink}), at large speeds, the kink and antikink are immediately reflected upon collision. When their speeds are below some critical speed, the kink and antikink annihilate through the formation of a chaotic radiating bound state. We have observed both effects (the reflection and annihilation) in our numerical experiments with buckled graphene.

\begin{figure}[t]%
\includegraphics[width=80mm]{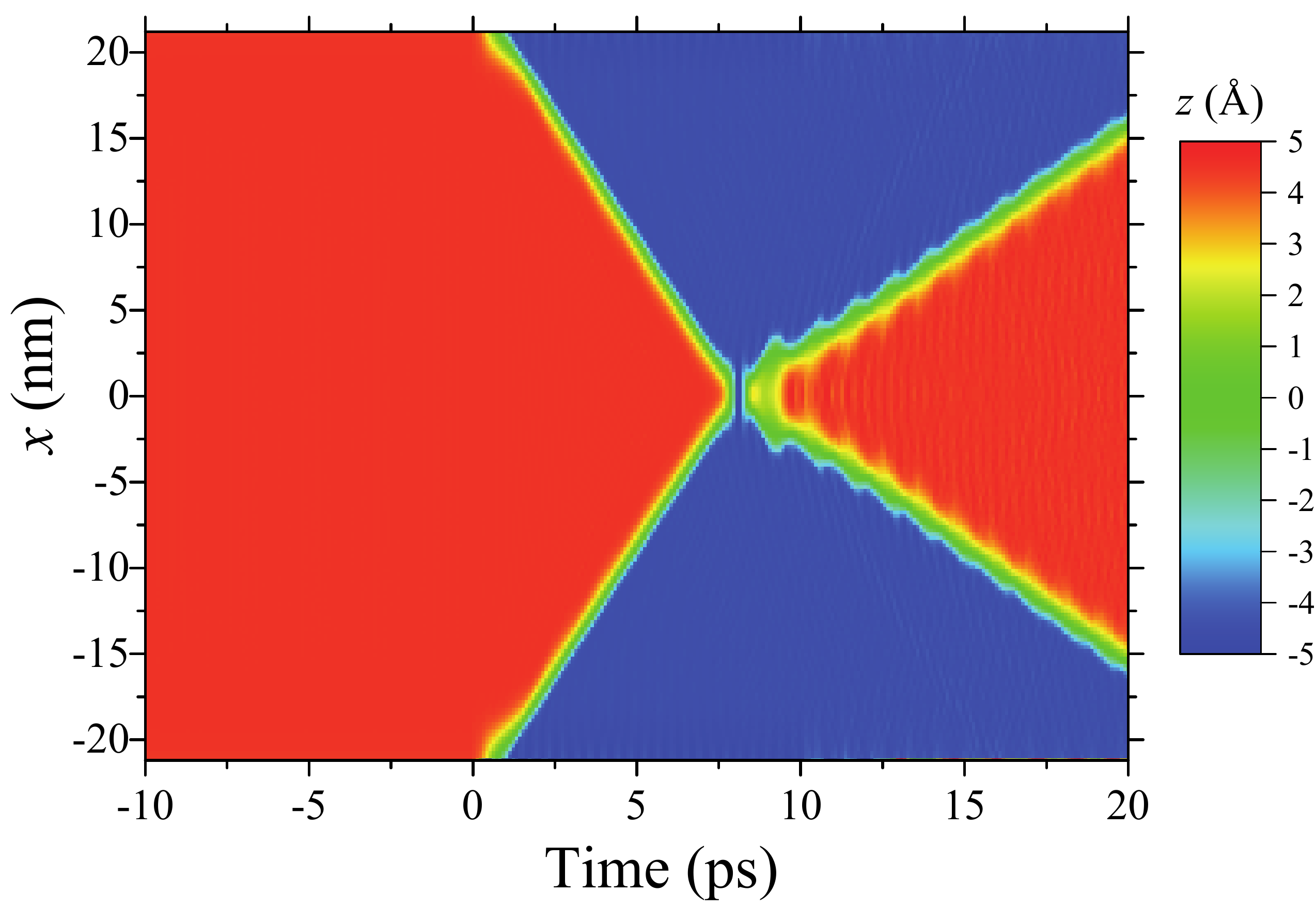}
\caption{A bounce resonance in kink-antikink scattering. This plot was obtained similarly to Fig. \ref{fig4} using the pulling force of 61 pN/atom.}
\label{fig5}
\end{figure}

In order to create slower moving graphene kinks and antikinks, we have slightly modified the above approach by changing the constant forces with linearly decreasing forces (from a maximum value at the edge to 0 in the bulk) applied to the same group of atoms. Fig. \ref{fig4} (a) demonstrates that
when the kink and antikink moving at the speeds of about 2.9 km/s  collide, they indeed form a bound state dissipating with time. The kink and antikink colliding at about 4.4 km/s are immediately reflected (see Fig. \ref{fig4} (b)). Moreover, Fig. \ref{fig5} presents a bounce resonance-like~\cite{goodman2005kink} behavior in the kink-antikink collision that is observed in the vicinity of the transition from annihilation to reflection collision regimes. Fig. \ref{fig4} (b) and Fig. \ref{fig5} clearly demonstrate that the graphene kink-antikink collisions are inelastic just as collisions of $\phi^4$ kinks/antikinks~\cite{campbell1983}.

In this Letter, we have investigated the properties of buckled graphene membranes above long but narrow trenches.
Using molecular dynamics simulations, we have explicitly demonstrated the existence of topologically non-trivial
solutions -- kinks and antikinks -- in such membranes and found some of their properties.
Our numerical simulations reveal a complex dynamics of stressed graphene nanoribbon, which is a 2D system of many atoms  with a large number of degrees of freedom.

It's interesting that in many respects the graphene kinks are similar to the kinks in the classical $\phi^4$ field model in 1+1 dimensions. In particular, both the buckled graphene and $\phi^4$ model have two stable equilibrium states and topologically non-trivial solutions connecting these states. In both cases, there is a limiting speed for the kink propagation, and the single kink (at $T=0$) propagates without loosing its energy. Additionally, in the buckled graphene the kink-antikink scattering shows some typical to $\phi^4$ field model features, such as the possibilities of kink-antikink annihilation and reflection from each other.

At the same time we have noticed some small differences such as a slightly assymetric shape of the graphene kink (Fig. \ref{fig2}(b)) and the existence of a non-symmetric kink, which might be considered as an excited state of the symmetric one. Overall, we believe that such differences are minor and postpone to future work the investigation of their role. Our main conclusion is that the buckled graphene membranes offer an interesting opportunity to test the predictions of $\phi^4$ model. We also expect that this system may find some unexpected technological applications.

This work has been partially supported by the Russian Science Foundation grant No. 15-13-20021.

\bibliography{memcap}

\end{document}